\documentclass[]{elsarticle}
\usepackage{enumerate}
\usepackage{dsfont}
\usepackage{amsfonts}
\usepackage{amssymb}
\usepackage[latin1]{inputenc}
\usepackage[T1]{fontenc}
\usepackage{graphicx}
\usepackage{mathtools}
\usepackage{bbold}	

\usepackage[dvipsnames]{xcolor}

\usepackage[colorlinks=true]{hyperref}
\usepackage{amsthm}

\biboptions{sort&compress}

\begin{document}

\begin{frontmatter}
\title{Redshift and lateshift from homogeneous and isotropic modified dispersion relations}

\author[UT]{Christian Pfeifer}
\ead{christian.pfeifer@ut.ee}
\address[UT]{Laboratory of Theoretical Physics, Institute of Physics, University of Tartu,\\ W. Ostwaldi~1, 50411 Tartu, Estonia}

\begin{abstract}
Observables which would indicate a modified vacuum dispersion relations, possibly caused by quantum gravity effects, are a four momentum dependence of the cosmological redshift and the existence of a so called lateshift effect for massless or very light particles. Existence or non-existence of the latter is currently analyzed on the basis of the available observational data from gamma-ray bursts and compared to predictions of specific modified dispersion relation models. We consider the most general perturbation of the general relativistic dispersion relation of freely falling particles on homogeneous and isotropic spacetimes and derive the red- and lateshift to first order in the perturbation. Our result generalizes the existing formulae in the literature and we find that there exist modified dispersion relations causing both, one or none of the two effects to first order.
\end{abstract}


\end{frontmatter}

\section{Introduction}
Most information about the properties of gravity are obtained by probing the geometry of spacetime through the observation of freely falling particles. In order to observe traces of the expected quantum nature of the gravitational interaction, one option is to look for their manifestation in the propagation of particles through spacetime, which we observe with telescopes. The theoretical prediction of such effects is one branch of quantum gravity phenomenology~\cite{Amelino-Camelia2013}. The pictorial idea why quantum gravity effects may become visible in this way is the following. Test particles probe spacetime on length scales which are inverse proportional to their energy. Thus the higher the energy of the particles, the smaller the length scale probed. Quantum gravity effects are expected to become relevant at the Planck scale and hence particles with energies closer to the Planck energy $E_{pl}$ should interact stronger with the quantum nature of gravity than lower energetic ones. Therefore, the propagation of high energetic particles through spacetime may deviate from their predicted behavior by classical general relativity. Since the energy of a particle is observer dependent this pictorial idea needs to be formulated more precisely in terms of the particle's four momentum, instead of its energy, what we will do during the derivations of this letter.

As long as a fundamental theory of quantum gravity is not available to predict this effect from the scattering between gravitons and the probe particles such quantum gravity effects can be modeled phenomenologically by a modification of the relativistic dispersion relation of freely falling point particles, see~\cite{Matschull:1997du,AmelinoCamelia:1999pm,KowalskiGlikman:2002jr,Freidel:2003sp,Myers:2003fd, Freidel:2005me,Ling:2005bq,Mattingly:2005re,Girelli:2006sc,Liberati:2013xla,Nilsson:2016rsv} and references therein.

Even though the particles we observer have energies below the Planck energy, the small effect may accumulate over a long particle travel time and become detectable. In particular observations from high redshift gamma-ray bursts (GRBs) are candidates to find traces of Planck scale induced modified dispersion relations (MDR) \cite{AmelinoCamelia:1997gz,Jacob:2006gn,AmelinoCamelia:2009pg,Stecker:2014oxa}. One most prominent signature would be a so called lateshift observation~\cite{Amelino-Camelia:2013uya}, i.e.\ an advance or a delay in the expected time of arrival of high energetic photons and neutrinos from the same source compared to low energetic ones emitted at the same time. Recently a preliminary analysis of the ICECUBE data for such a lateshift has been performed in \cite{Amelino-Camelia:2016ohi} as well as an analysis of GRBs detected with the Fermi Gamma-Ray Space Telescope~\cite{Xu:2018ien,Xu:2016zsa,Xu:2016zxi}.

To deduce a MDR from the measured time of arrival data of neutrinos and photons from GRBs a derivation of the lateshift effect from a most general modification of the general relativistic dispersion relation is required. Usually specific models are assumed and the lateshift is derived for these classes of MDRs~\cite{Amelino-Camelia2013,Jacob:2006gn,AmelinoCamelia:2009pg,Ellis:2002in,Ellis:2005wr,Rosati:2015pga}. 

In this letter we derive the redshift and lateshift from an arbitrary perturbation of the general relativistic dispersion relation to first order in the perturbation. Observation or not-observation of a modified redshift or a lateshift effect then directly leads to conditions the perturbation of the dispersion relation must satisfy to be viable. As an interesting insight from the general red- and lateshift formula we findMDRs which predict both aforementioned effects, only one of them or even none to first order.

\section{Dispersion relations as Hamilton functions on spacetime}
To derive the lateshift from the dispersion relation of point particles on spacetime we interpret a dispersion relation as level sets of a Hamilton function on the spacetime's cotangent bundle, as it turned out to be a very useful framework to treat MDRs on curved spacetimes covariantly~\cite{Barcaroli:2015xda,Barcaroli:2016yrl,Barcaroli:2017gvg}.

The four momentum of a particle is a $1$-form $P$ on spacetime which can be expanded in local coordinates around a point $x$ as $P=p_a dx^a$. The tuple $(x,p)$ denotes the particle's momentum~$p$ at the spacetime position~$x$. A dispersion relation is a level set of a Hamilton function $H(x,p)$ which determines the particle's motion. This covariant formulation of dispersion relations on curved spacetimes has the advantage that it allows to study dispersion relations on the basis of the particle's four momentum without referring to the observer dependent notion of a particle's energy or spatial momentum.

Homogeneous and isotropic dispersion relations are characterized by Hamilton functions with a specific dependence on the particle's positions and momenta. As shown in \cite{Barcaroli:2016yrl} the most general homogeneous and isotropic dispersion relation is given by the level sets of the Hamiltonian
\begin{align}
	H(x,p) = H(t,p_t, w),\quad w^2 = p_r^2 \chi^2 + \frac{p_\theta^2}{r^2} + \frac{p_\phi^2}{r^2 \sin^2\theta}\,,
\end{align} 
where $\chi = \sqrt{1-k r^2}$. Due to the high symmetry the Hamilton equations of motion, which determine the propagation of the particle through spacetime, can partly be solved and reduce to 

\begin{minipage}{0.45\textwidth} 
	\begin{align}
	\dot p_t&=- \partial_t H\,,\label{eq:dotpt}  \\
    p_r&=\frac{K_1}{\chi} \label{eq:pr} \,, \\
	p_\theta &= 0\,,\\
	p_\phi&=0\,,
	\end{align}
\end{minipage}
\hfill
\begin{minipage}{0.45\textwidth}
	\begin{align}
	\dot t &= \partial_{p_t} H\, \label{eq:dott}, \\
	\dot r & = \partial_wH \frac{1}{w} \chi K_1\, \label{eq:dotr},\\
	\theta & = \frac{\pi}{2}\,,\\
	\phi &= 0\,,
	\end{align}
\end{minipage}\\

\noindent where $K_1^2=w^2$ is a constant of motion.

\section{The perturbed dispersion relation}
The most general perturbation of the homogeneous and isotropic general relativistic dispersion relation is given by the level sets of
\begin{align}\label{eq:H}
	H(t,p_t,w) = - p_t{}^2 + a(t)^{-2} w^2 + \epsilon h(t,p_t,w)\,.
\end{align}
The perturbation $h(t,p_t,w)$ can be an arbitrary function of $t$, $p_t$ and $w$, and $\epsilon$ is an arbitrary perturbation parameter. In the context of quantum gravity or Planck scale induced perturbations it may be identified with the Planck scale, while other sources of a modification of the dispersion relation may require a different perturbation parameter. For the calculations below we do not fix the origin of the perturbation.

To derive the redshift and lateshift from \eqref{eq:H} we use the Hamilton equations of motion
\begin{align}\label{eq:tdotrdot}
	\dot t = -2 p_t + \epsilon \partial_{p_t} h\,,\quad \dot r = \chi \bigg( \frac{2w}{a^2} + \epsilon \partial_w h\bigg)\,,
\end{align}
and the dispersion relation
\begin{align}\label{eq:disprel}
	- p_t{}^2 + a^{-2} w^2 + \epsilon h(t,p_t,w)  = - m^2\,.
\end{align} 
The time dependence of the scale factor $a$ will from now on only be displayed when necessary.

\subsection{Redshift}
The dispersion relation \eqref{eq:disprel} determines $p_t$ as function of $t,r$ and $w$ without solving any equation of motion. From the ansatz $p_t = p_t^0 + \epsilon p_t^1$ one easily finds
\begin{align}\label{eq:ptmass}
	p_t(t,w,m) = - \sqrt{m^2 + \frac{w^2}{a^2}} + \epsilon \frac{h(t,p_t^0(t,w,m),w)}{2 p_t^0(t,w,m)}\,,
\end{align}
and thus for massless particles
\begin{align}\label{eq:ptmassless}
	p_t(t,w,0) = - \frac{w}{a} - \epsilon \frac{a}{2w}h(t, p_t^0(t,w,0), w)\,.
\end{align}
The redshift of a photon which is emitted at time $t_i$ with a coordinate time-momentum $p_t(t_i,w) = p_t(t_i,w,0)$ and observed at time $t_f$ with coordinate momentum $p_t(t_f,w)= p_t(t_f,w,0)$, subject to the dispersion relation in consideration then is
\begin{align}\label{eq:redshift}
	&z(t_i, t_f) 
	= \frac{p_t(t_i,w)}{p_t(t_f,w)} - 1 \nonumber\\
	&= \bigg(\frac{a(t_f)}{a(t_i)}-1\bigg) - \frac{\epsilon}{2 w^2} \frac{a(t_f)}{a(t_i)} \bigg( a(t_f)^2 h(t_f,p_t^0(t_f,w),w) - a(t_i)^2 h(t_i,p_t^0(t_i,w),w)\bigg)\,.
\end{align}
To zeroth order, as expected, the redshift formula from general relativity is recovered, while the first order is determined by the perturbation $h$. In particular the perturbation depends in general on the particles spatial coordinate momentum $w$, which can be expressed in terms of the initial coordinate time-momentum of the photon $p_t(t_i)$, since equation \eqref{eq:ptmassless} can be inverted for $w(p_t,t)$. Thus photons starting with different initial coordinate time-momentum $p^0_t(t_i, w)$ experience a different redshift. Hence a detection of a photon redshift dependent on the initial coordinate time-momentum is a clear signal for a modification of the dispersion relation while its absence puts constraints on the perturbation. First analyses of possible evidences for an energy dependent redshift have been performed \cite{Arai:2016bim,Ferreras:2016xsq}.

We use the term coordinate time-momentum of a photon here instead of energy of a photon to distinguish between the observer dependent notion of energy of a particle and the observer independent choice of coordinates to describe the particle's four momentum.
 
\subsection{Lateshift}\label{sec:latehishift}
To derive the lateshift we need to solve the radial equation of motion \eqref{eq:dotr}, which is done best when $r$ is parametrized in terms of the coordinate time by using \eqref{eq:tdotrdot}
\begin{align}\label{eq:rprime}
	\frac{\mathrm{d}r}{\mathrm{d}t}= \frac{\dot r}{\dot t}
	&= \frac{\chi w}{a\sqrt{a^2\ m^2+w^2}}\bigg(1 - \epsilon \frac{1}{2 (p_t^0)^2} \bigg[ h(t,p_t^0,w) - p_t^0 \partial_{p_t} h(t,p_t^0,w) - w \partial_w h(t,p_t^0,w)  \bigg]\bigg)\nonumber\\
	&\equiv \frac{\chi w}{a\sqrt{a^2\ m^2+w^2}}(1 - \epsilon f(t, p_t^0, w))\,.
\end{align}
The momentum corresponding to the time coordinate is considered as function $p_t = p_t(t,w,m)$ as displayed in \eqref{eq:ptmass}.
Employing separation of variables and the perturbative ansatz  $r = r^0 + \epsilon r^1$ the following solution can easily be found
\begin{align}\label{eq:rtwm}
	r(t,w,m) 
	&= \frac{1}{\sqrt{k}}\sin\bigg(\sqrt{k} C + \sqrt{k} \int_{t_i}^{t} \mathrm{d}\tau\ \frac{ w}{a\sqrt{a^2\ m^2+w^2}}\bigg) \nonumber\\
	&- \epsilon \cos\bigg(\sqrt{k} C + \sqrt{k} \int_{t_i}^{t} \mathrm{d}\tau\ \frac{ w}{a\sqrt{a^2\ m^2+w^2}}\bigg)\ \int_{t_i}^{t}\mathrm{d}\tau\ \frac{w\ f(\tau,p_t^0,w)}{a\sqrt{a^2\ m^2+w^2}}\,.
\end{align}
Observe, that for massive particles even the zeroth order depends on the spatial momentum $w$, respectively on the particles initial coordinate time-momentum in case one considers the spatial momentum as function of the initial momentum by solving equation \eqref{eq:ptmass} for $w(p_t, t)$. For massless particles this dependence vanishes and only appears in the first order correction.

The search for lateshift effects focuses on neutrinos and photons, i.e. particles of light or zero mass \cite{Amelino-Camelia:2016ohi,Xu:2018ien}. To derive the lateshift for both we expand \eqref{eq:rtwm} for small masses and find the  first non-vanishing order, neglecting the order $\epsilon m^2$ and higher orders in $\epsilon$ or $m^2$,
\begin{align}
	r(t,w,m)
	&= \frac{1}{\sqrt{k}}\sin\bigg(\sqrt{k} C + \sqrt{k} \int_{t_i}^{t} \mathrm{d}\tau\ \frac{1}{a}\bigg) - m^2 \cos\bigg(\sqrt{k} C + \sqrt{k} \int_{t_i}^{t} \mathrm{d}\tau\ \frac{1}{a}\bigg) \int_{t_i}^t \mathrm{d}\tau\ \frac{a}{2 w}\nonumber\\
	&- \epsilon \cos\bigg(\sqrt{k} C + \sqrt{k} \int_{t_i}^{t} \mathrm{d}\tau\ \frac{1}{a}\bigg)\int_{t_i}^{t}\mathrm{d}\tau\ \frac{ f(\tau,p_t^0,w)}{a}\bigg|_{p_t^0=p_t^0(t,w,0)} + \mathcal{O}(\epsilon^2, \epsilon m^2, m^4) \,.
\end{align}

Consider two radially freely falling particles of in general different masses $m_1$ and $m_2$ of same order, with different momenta $w_1$ and $w_2$. They shall be emitted at the same initial time $t_i$ at the origin of the coordinate system. We call their trajectories $r(t_1, w_1,  m_1)$ and $r(t_2, w_2, m_2)$ respectively. Thus the condition that they reach the same radial coordinate distance $R$ in spacetime is $r(t_1, w_1,m_1) = r(t_2, w_2,m_2)$. 

Introducing the mass lateshift $\Delta t_{m}$ and the lateshift due to the MDR $\Delta t_\epsilon$ we make the ansatz $t_2 = t_1 + \alpha \Delta t_m + \epsilon \Delta t_\epsilon$ for the time of arrival of the second particle at $R$, where $\alpha$ is an order parameter which counts the order of the masses. Solving the equal position condition order by order yields the lateshift formulas
\begin{equation}
	\Delta t_m = \frac{m^2_2 w_1 - m^2_1 w_2 }{2w_1w_2} a(t_1) \int_{t_i}^{t_1}\mathrm{d}\tau\ a(\tau)
\end{equation}
and
\begin{align}\label{eq:lateshift}
	\Delta t_\epsilon = a(t_1) \int_{t_i}^{t_1}\mathrm{d}\tau\ \frac{f(\tau,p_t^0(\tau,w_2),w_2) - f(\tau,p_t^0(\tau,w_1),w_1)}{a(\tau)}\,.
\end{align}
Since $f(t,p_t^0,w)$ may depend arbitrarily on $t$ and not only through $a(t)$ it is in general not possible to rewrite this equation in terms of the zeroth order redshift of the particles $z(t)=z(t,t_f)$ at their emission time. Only if $z(t)$ is solvable for $t$ or, if the time dependence of the perturbation $f(t,p_t^0,w)$ can be expressed as a function of the scale factor $a(t)$, one may express the lateshift in terms of a redshift integral, the Hubble parameter $\mathcal{H}$ and the cosmological density parameters $\Omega_\Lambda$, $\Omega_k$ and~$\Omega_M$
\begin{align}
\begin{split}
	\Delta t_\epsilon 
	&= \int_{0}^{z}\mathrm{d}z'\ \frac{f(z',w_2) - f(z',w_1)}{\mathcal{H}(z')} \\
	&= \int_{0}^{z}\mathrm{d}z'\ \frac{f(z',w_2) - f(z',w_1)}{\mathcal{H}(0)\sqrt{\Omega_\Lambda + \Omega_k(1+z)^2 + \Omega_M(1+z)^3}}\,.
\end{split}
\end{align}
In this form the lateshift derived from the general dispersion relation~\eqref{eq:disprel} can be recognized as generalization of the expression derived in \cite{Ellis:2002in,Ellis:2005wr}, which is employed in the data analyses \cite{Amelino-Camelia:2016ohi,Xu:2018ien,Xu:2016zsa,Xu:2016zxi}.

For the derivation of the lateshift effect we assumed the simultaneous emission of particles with different momenta here, to demonstrate how the effect is predicted from the MDR \eqref{eq:H}. In GRBs this assumption is not necessarily realized and one has to take into account that the observed lateshift $\Delta t_{obs} = \Delta t_m + \Delta t_\epsilon + \Delta t_{int}$ is composed of an arrival delay due to the particles mass $\Delta t_m$, the lateshift caused by the MDR $\Delta t_\epsilon$, and, in addition, a difference in the emission time of particle of different momentum due to the mechanism of the GRB itself~$\Delta t_{int}$~\cite{CHANG201247}. The latter may be derived from a fundamental model of the GRB and must be subtracted from the observed value to identify the lateshift effect due to the MDR which we discussed. In the future it may be possible to derive an additional modification of the MDR \eqref{eq:H} from a GRB model which implements a difference in the emission time of particles of different energies directly in the calculation done here.

We conclude that the measurement of a lateshift effect for massless particles which are emitted at the same time would be a clear indication of a MDR. A detection of a lateshift effect for light massive particles due to a MDR may be more difficult to identify due to the additional effect coming from the mass lateshift. For all kinds of detections of a lateshift effect it is necessary to analyse possible uncertainties in the emission time of the particles of different momenta.

\section{Examples}
The general first order redshift and lateshift formulae \eqref{eq:redshift} and \eqref{eq:lateshift} enable us to determine if a MDR yields an energy dependent redshift and a lateshift, only one of both effects or none. In the following we give examples for each case.

For perturbations $h(t,p_t,w)$ which are homogeneous of degree $r$ in the variables $p_t$ and $w$, i.e.\ which satisfy $h(t, \lambda p_t, \lambda w) = \lambda^r h(t,p_t,w)$, the function $f(t,p_t^0,w)$ which causes the lateshift \eqref{eq:lateshift} simplifies to
\begin{align}
f(t,p_t^0,w) = \frac{1}{2(p_t^0)^2} (1-r)h(t,p^0_t,w) = \frac{w^r}{2(p_t^0)^2} (1-r)h(t,\tfrac{p^0_t}{w},1) \,,
\end{align}
by its definition in \eqref{eq:rprime} and Euler's theorem for homogeneous functions.
Calculating the lateshift for a generic third order polynomial $h = b(t) p_t{}^3 + c(t) p_t{}^2 w+ d(t) p_t w^2 + e(t) w^3$, i.e.\ $r=3$, thus yields
\begin{align}\label{eq:latepoly}
\Delta t_\epsilon
= \frac{a(t_1)(w_1-w_2)}{2} \int_{t_i}^{t_1}\mathrm{d}\tau\ \frac{1}{a(\tau)} \bigg(a(\tau)^2 e(\tau) - a(\tau) d(\tau) + c(\tau)- \frac{ b(\tau)}{a(\tau)}\bigg)\,,
\end{align}
while the first order perturbation in the redshift becomes linear in $w$
\begin{align}\label{eq:redpoly}
\begin{split}
&1-(z(t_i,t_1) +1)\frac{a(t_i)}{a(t_1)}\\
&= \frac{\epsilon}{2}w\bigg(a(t_1)^2 e(t_1) - a(t_1) d(t_1) + c(t_1) - \frac{b(t_1)}{a(t_1)} - a(t_i)^2 e(t_i) + a(t_i) d(t_i) - c(t_i) + \frac{b(t_i)}{a(t_i)}\bigg)\,.
\end{split}
\end{align}
Hence in particular if the integrand in \eqref{eq:latepoly} vanishes, which means $h(t, p_t^0, w) = 0$ in this case, the lateshift of the MDR vanishes and the redshift is as on Friedmann-Lema\^{i}tre-Robertson-Walker spacetimes, independent of the particles four momentum. Other examples for perturbations which share this properties are $h = (- p_t{}^2 + a(t)^{-2}w^2)Q(t,p_t,w)$ for arbitrary $Q(t,p_t,w)$.

Another class of perturbations, the ones of the form $h = p_t{}^n Q_t(t, X) + w^n Q_w(t, X)$ with $X=\frac{w}{p_t}$, $Q_t$ and $Q_w$ being arbitrary functions of their arguments and $n=1,2$, do not induce a lateshift. For $n=1$ the redshift becomes four momentum dependent while for $n=2$ it only picks up a four momentum independent correction.

In the context of quantum gravity phenomenology a most intensively studied MDR is the $\kappa$-Poincar\'e dispersion relation \cite{Gubitosi:2013rna,KowalskiGlikman:2001px}. To first order in the Planck length it is of the polynomial type discussed above with $b(t)=c(t)=e(t)=0$ and $d(t)=a(t)^{-2}$. Employing this identification in \eqref{eq:latepoly} and \eqref{eq:redpoly} reproduces the lateshift and redshift results known in the literature \cite{Rosati:2015pga,Barcaroli:2016yrl,Barcaroli:2015eqe}.

\section{Beyond homogeneous and isotropic dispersion relations}
In this letter we considered perturbations of the homogeneous and isotropic general relativistic dispersion relation, which are themselves again homogeneous and isotropic. For upcoming studies we aim to investigate the observable effects of more general perturbations. The necessary change to do so is to consider general perturbation functions $h(x,p)$ in the Hamiltonian \eqref{eq:H} and not only those which depend on $(t,p_t,w(r,\theta, \phi, p_r, p_\theta, p_\phi))$. How to treat general non-homogeneous modified dispersion relations in terms of Hamilton functions on curved spacetime has been developed in \cite{Barcaroli:2015xda,Barcaroli:2017gvg}.
 
A general ansatz for a perturbation can for example be expressed as power series in the momenta
\begin{align}
h(x,p) = \sum_{i=0}^\infty f^{a_1a_2...a_i}(x)p_{a_1} p_{a_2} ... p_{a_N}\,,
\end{align}
where the coefficient functions $ f^{a_1a_2...a_i}(x)$ specify the support and the type of the perturbation. Depending on the phenomenon one seeks to describe in terms of MDRs these functions may have different origins.

In the context of GRBs such terms may be added to the homogeneous and isotropic one in \eqref{eq:H} to describe the motion of particles derived from a fundamental GRB model, as already mentioned at the end of section \ref{sec:latehishift}.

Further interesting models to investigate are MDRs which depend on the local standard model matter and dark matter distribution on spacetime as well as the zeroth order redshift. In the context of string theory  \cite{Ellis:2008gg} as well as in the study of the interaction of light with the matter content of the universe \cite{Latimer:2013rja} such dispersion relations emerge.

One way to realize such models is the following. Let $z$ be the zeroth order redshift, $\rho_{SM}(x)$ be the matter density of standard model particles in the universe and $\rho_{DM}(x)$ be the dark matter density in the universe. A general power law model which realizes a MDR depending on these quantities would be given by the functions
\begin{align}
	f^{a_1a_2...a_i}(x) = c^{a_1a_2...a_i} \sum_{j=-\infty}^\infty \sum_{k=-\infty}^\infty \sum_{l=-\infty}^\infty C_{jkl}\ z^j\ \rho_{SM}(x)^k\ \rho_{DM}(x)^l
\end{align}
for constants $c^{a_1a_2...a_i}$ and $C_{jkl}$ whose value can either be predicted by fundamental theories which cause the MDR or be obtained from observations.

The algorithm we outlined here is not restricted to study observable consequences of MDRs in cosmology but can be applied to any spacetime of interest by starting instead of from \eqref{eq:H} from a general perturbation of a metric Hamiltonian
\begin{align}
	H(x,p) = g^{ab}(x)p_ap_b + \epsilon h(x,p)\,.
\end{align}

\section{Conclusion}
Starting from a general first order perturbation of the general relativistic homogeneous and isotropic dispersion relation of freely falling point particles~\eqref{eq:H} we derived the observables redshift~\eqref{eq:redshift} and lateshift \eqref{eq:lateshift}. Compared to general relativity the redshift generically becomes energy dependent and the lateshift for simultaneously emitted photons emerges. With help of the new general first order formulae obtained here it was possible to demonstrate that there exist particular MDRs in which only one or none of the effects appear. 

Observation or non-observations of a four momentum dependent redshift or lateshift of particles from the same source emitted at the same time now directly leads to bounds, which the first order perturbation of the dispersion relation must satisfy. The interpretation of a lateshift observation must however be done with care due to uncertainties in the simultaneity of the emission time of the particles. To identify the effects coming from a quantum gravity induced MDR it is necessary to take such emission delays into account.

For future studies the methods we applied here to study MDRs on curved spacetimes can be applied to systematically compare observational results with predictions from MDRs, not only in the context of cosmology but for all kinds of physical systems of interest such as for example black hole and gravitational wave spacetimes.\\

\textbf{Acknowledgments:}
Thanks to Giulia Gubitosi, Mihkel R\"unkla and Manuel Hohmann for very helpful comments and remarks. C.P.\ gratefully thanks the European Regional Development Fund through the Center of Excellence TK133 ``The Dark Side of the Universe'' for financial support.

\bibliography{MIMDRCosmo}

\end{document}